\documentclass[11pt]{article}
\usepackage{moriond,epsfig}

\def\be{\begin{equation}}
\def\ee{\end{equation}}
\def\bea{\begin{eqnarray}}
\def\eea{\end{eqnarray}}

\begin{document}
\vspace*{4cm}
\title{Cluster Mergers and Diffuse Radio Emission in Abell 2256 and Abell 754}

\author{ T.\ E.\ Clarke }

\address{National Radio Astronomy Observatory, 1003 Lopezville Dr., Socorro, NM, 87801, USA}

\author{ T.\ A.\ Ensslin}

\address{Max-Planck-Institut f\"ur Astrophysik, Karl-Schwarzschild-Str.\ 1, 85741 Garching, Germany}

\maketitle
\abstracts{ We present deep VLA observations of the galaxy
clusters Abell 2256 and Abell 754, both of which appear to be in the
violent stage of major cluster merger events. The complex nature of
Abell 2256 is revealed through radio images which, in addition to the
head-tail galaxies, show two extended, irregular, and sharp-edged
regions of diffuse radio emission at the cluster periphery (so called
radio relics), and a large-scale diffuse radio halo located in the
central regions of the cluster. Polarimetry of the A2256 cluster
relics reveals large-scale ordered magnetic fields which appear to
trace the bright filaments in the relics. The polarization fraction
across the relics ranges from 20\% - 40\% with the majority of the
relics polarized above the 30\% level. At the sensitivity of our
current observations we place an upper limit of 20\% on the
polarization of the radio halo. Low frequency VLA observations of
Abell 754 reveal extended, diffuse radio (halo) emission in the
cluster core region as well as steep spectrum emission in the cluster
periphery. The location, morphology, and spectral index of the
peripheral emission are consistent with the properties of radio
relics. The X-ray evidence of the ongoing mergers in both clusters,
together with the polarization properties of A2256's radio relics
supports recent suggestions of a merger-induced origin of the relic
emission. Deciphering the complex radio properties of these clusters
may thus provide the key to understanding the dynamical history of the
systems. }

\section{Introduction}

In the hierarchical model of structure formation objects form from the
collapse of initial density enhancements and subsequently grow through
gravitational effects. Numerical simulation of structure formation
show that clusters of galaxies are found to preferentially form at the
intersection points of large filamentary structures
\cite{w91,kw93}. Due to the high-density environment in which they
reside, clusters of galaxies are thus expected to undergo several
merger events as they form. These merger events are highly energetic
($10^{63}-10^{64}$ ergs) and thus provide a significant energy input
into the intracluster medium (ICM). Large scale structure simulations
\cite{m00} as well as recent 3D MHD/N-body simulations
\cite{r99a,r99b} find that the shocks and turbulence associated with a
major cluster merger event can significantly amplify the intracluster
magnetic field and accelerate relativistic particles which, in the
presence of a magnetic field, emit synchrotron emission.

\subsection{Diffuse Radio Emission in Galaxy Clusters}

Radio observations toward a number of galaxy clusters reveals
the presence of large regions of diffuse radio emission which extend
over scales of $>$ 600 kpc in the ICM and have no obvious optical
counterpart. This emission appears to fall in two categories: {\it
halos} which are centrally located in the cluster, relatively regular
in shape, and unpolarized, and {\it relics} which are peripherally
located, fairly elongated and irregular, and often highly
polarized \cite{fg96}.

The presence of these large regions of diffuse synchrotron emission
reveals the large scale distribution of relativistic particles and
magnetic fields in the intracluster medium. Despite extensive searches
through the NRAO VLA Sky Survey \cite{gtf} and the Westerbork Northern
Sky Survey \cite{ks} as well as a number of targeted searches for
radio halos and
relics \cite{gf00,vbm,rhlp,sr,fbgn,hswd,c85,hsj,lhba,kcecn} there are
still a relatively small number of clusters known to contain this
emission. We must therefore ask: Why are these sources so rare? Do the
known clusters contain excess magnetic fields? Do they contain a
larger reservoir of relativistic particles than other clusters? or
perhaps they have excess particles and magnetic fields?

Many, if not all, of the galaxy clusters which are confirmed to
contain diffuse radio emission also show significant evidence of
merger activity. The clusters which contain radio halos tend to be
very X-ray luminous, massive clusters \cite{cola,lhba,buote} which
display a significant amount of X-ray substructure. Although this
suggests that the merger event is the trigger for the diffuse
emission, it should be noted that only 10\% of galaxy clusters appear
to contain diffuse emission while more than 40\% of clusters show
evidence of merger activity \cite{jf}. 
 
\section{Abell 2256, a Mpc$^3$ Non-Thermal Laboratory}

\subsection{Two or Three Body Merger System?}

Abell 2256 is a rich, nearby (z=0.0594) cluster of galaxies and was
one of the first targets observed by ROSAT. Analysis of these
observations \cite{b91a} revealed significant substructure in the
cluster. The X-ray surface brightness distribution shows two separate
X-ray peaks that indicate it is undergoing a merger event.  The X-ray
temperature map of the cluster \cite{b94} shows that the infalling
component is cooler than the main cluster body and that there are two
hot regions roughly perpendicular to the merger axis. These hot
regions are similar to those seen in simulations of merger events
where the merger has not yet proceeded past core
passage \cite{sm,rlb}. The ROSAT view of the cluster is thus a two body
merger which is approximately 1 Gyr before core passage. 

Although the two body merger model fits the X-ray data quite well, it
will be seen below that the radio data are very hard to understand in
this scenario. A couple of days before the Moriond meeting, a Chandra
paper on Abell 2256 appeared on {\it astro-ph} \cite{sun} which showed a new
structure was visible in the X-ray surface brightness map in addition
to the merging component found by ROSAT. This feature is located close
to the core of the cluster and is interpreted as either another
merging component, or an internal structure to the main cluster. If
the new feature found by Chandra is interpreted as a merger signature,
comparison of the the X-ray images to cluster merger
simulations \cite{t99,t00} suggests that is may be an old merger which
is $\sim$ 0.3 Gyr after core passage. In this scenario, Abell 2256 is
a three-body merger system where the oldest merger has proceeded
significantly far as to only be visible as an X-ray excess near the
cluster core, while the current merger is still in the early stages
and has not significantly disrupted the cluster gas. This merger
history fits much better with the observed radio properties of the
cluster.

Further X-ray observations of the A2256 cluster with BeppoSAX reveal
the presence of a hard X-ray tail. Such emission has been detected in
Coma \cite{ff99} and Abell 2199 \cite{k99} and is generally interpreted
as non-thermal inverse Compton (IC) emission from
relativistic electrons scattering off the cosmic microwave background
(CMB) photons. Such a model requires that there be a large reservoir of
relativistic electrons in the intracluster medium. In the presence of
a magnetic field, these electrons will radiate synchrotron emission
and should be visible in the radio regime.

\subsection{Radio Emission}

Previous radio observations of Abell 2256 \cite{bf,rsm} have revealed
that it has very complex radio emission. The cluster contains a number
of head-tail galaxies, two large regions of diffuse radio relic
emission, and also possibly a central radio halo. In order to obtain a
better understanding of the complex radio emission in this cluster, we
have undertaken deep NRAO VLA\footnote{The National Radio Astronomy
Observatory is a facility of the National Science Foundation operated
under cooperative agreement by Associated Universities, Inc.}
observations in several configurations of Abell 2256 at a number of
different frequencies ranging from 4.8 GHz to 330 MHz. In this paper
we discuss the results from observations at 1.4 GHz in the VLA's C and
D configurations.

\begin{figure}
\begin{center}
%\rule{5cm}{0.2mm}\hfill\rule{5cm}{0.2mm}
%\vskip 2.5cm
\psfig{figure=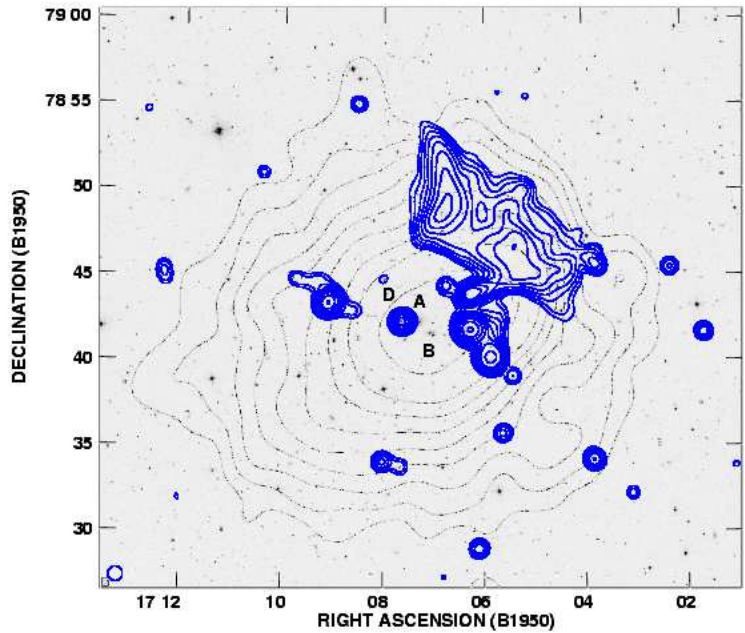,height=3.5in}
%\rule{5cm}{0.2mm}\hfill\rule{5cm}{0.2mm}
\caption{Digital Sky Survey R-band greyscale image overlaid with
smoothed ROSAT PSPC contours (black) and VLA 1.4 GHz contours
(blue). The galaxy notation in the figure is that of Sun et al.\
(2001). At the redshift of A2256, the linear scale is $\sim$ 1.1 kpc/arcsec for ${\rm H_o=75\ km\ s^{-1}\ Mpc^{-1}}$.
\label{fig_vla_nvss}}
\end{center}
\end{figure}

In Figure~\ref{fig_vla_nvss} the smoothed ROSAT X-ray contours of
A2256 are plotted over the R-band image from the Digitized Sky Survey
(DSS). Also shown on the image are the VLA 1.4 GHz radio contours
which reveal the compact and head-tail sources as well as the radio
relics. This ROSAT overlay can be compared to Figure 1 of Sun et al.\
(2001) which shows the DSS plus radio and Chandra contours. There are
three bright optical galaxies seen at the cluster center: galaxy 'D'
in the notation of Sun et al.\ (2001) which corresponds to the bright
compact radio source in the center of the image, the cluster center
elliptical ('B') which is located $\sim$ 2' south-west, has no radio
counterpart, and is only 0'.5 away from the central X-ray peak, and
galaxy 'A' which has a very extended optical halo but has not obvious
radio counterpart.

\begin{figure}
\begin{center}
%\rule{5cm}{0.2mm}\hfill\rule{5cm}{0.2mm}
%\vskip 2.5cm
\psfig{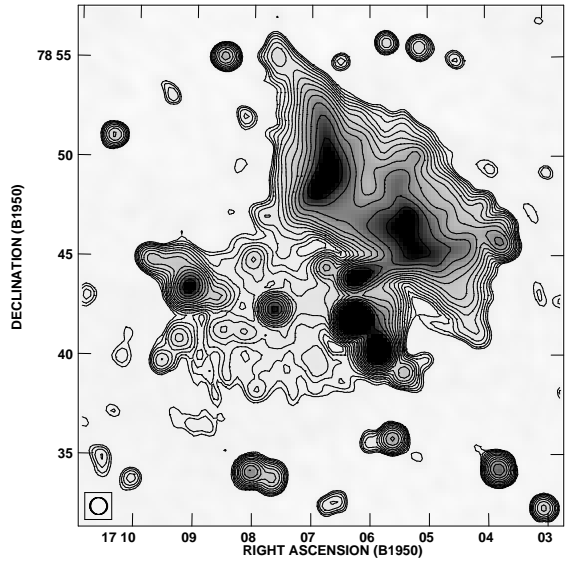}
%\rule{5cm}{0.2mm}\hfill\rule{5cm}{0.2mm}
\caption{VLA 1369 MHz D configuration total intensity greyscale and
contours. Both the peripheral radio relics and central radio halo
emission are clearly visible. 
\label{fig_vla_1369}}
\end{center}
\end{figure}

We have undertaken deep VLA observations of the complex radio emission
in Abell 2256 at four frequencies around 20 cm and two frequencies
around 6 cm with the VLA. These observations were undertaken in the
compact C and D configurations in order to provide the sensitivity
required to detect the diffuse, large-scale emission typical of
cluster halos. Figure~\ref{fig_vla_1369} shows the 1369 MHz VLA D
configuration total intensity image. The compact and head-tail sources
in the cluster are clearly visible along with the large regions of
diffuse radio emission. The elongated bright region to the north-west
is identified as radio relics G (the Eastern relic) and H (the
Western relic) in the notation of Bridle \& Fomalont (1976). The
image also reveals a large central halo in the cluster which was
previously marginally detected \cite{bf} but clearly extends to over
300 kpc in radius.

\begin{figure}
\begin{center}
%\rule{5cm}{0.2mm}\hfill\rule{5cm}{0.2mm}
%\vskip 2.5cm
\psfig{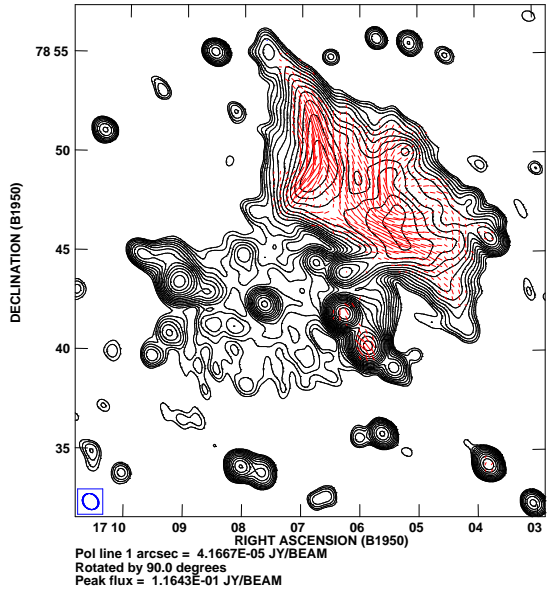}
%\rule{5cm}{0.2mm}\hfill\rule{5cm}{0.2mm}
\caption{VLA 1369 MHz D configuration contours of Abell 2256 with the
Faraday corrected magnetic field vectors overlaid.
\label{fig_vla_fdir}}
\end{center}
\end{figure}

The VLA observations show that the radio relics extend over a region
of 975 kpc $\times$ 650 kpc and appear to have a sharp edge to the
emission. Embedded with the relic structures are relatively bright
filamentary structures which appear to have widths of around 100 kpc
in the brighter G relic. The relics are highly polarized with the
linear polarization fraction above 30\% for the majority of the region
and reaching values of up to 50\%. The intrinsic magnetic field
direction shown in Figure~\ref{fig_vla_fdir} reveals that there is
large scale order to the fields, and it appears to trace the bright
filaments in the relics. The spectral index across the G relic is
remarkable uniform at -1.0 (${\rm S_{\nu} \propto \nu^{\alpha}}$)
between 20 cm and 6 cm. This spectral index is steeper than the value
of -0.37 found between 90 cm and 20 cm by Rottgering et al.\ (1994)
but is in good agreement with the more recent \cite{ks} value of -1.25
reported between 90 cm and 20 cm.

The radio halo appears to be offset from the cluster core, and is
centered roughly around a compact radio source which corresponds to
galaxy 'D' in Figure~\ref{fig_vla_nvss}. The halo emission is very
uniform, relatively symmetric, and shows no obvious edge. At the
sensitivity of our current observations, we place an upper limit of
20\% on the linear polarization of the halo. The halo appears to be
very steep spectral index emission, and we estimate a spectral index
of $\sim$ -2.0 between 20 cm and 6 cm.

The position of the emission for the radio relics is roughly
co-incident with the location of the merging subcluster to the
north-west of the cluster core. The radio halo on the other hand
appears to be centered about the current location of the old merger
remnant to the west of the cluster core. The relatively flat spectral
index of the relic emission suggests that the relativistic particles
may have recently been injected in the region or have undergone an
acceleration process. This is consistent with the picture that the
on-going merger event has recently shock-accelerated the particle
population in the north-west quadrant of the
cluster \cite{ebkk98,eg01}. The high polarization fraction and
alignment of the magnetic fields are consistent with shock compression
of an old magnetized plasma \cite{ebkk98,eb01}. The steeper
spectral index of the halo indicates that it has not recently received
a fresh population of relativistic particles. This is consistent with
the scenario of the halo formation from the earlier merger event.

\section{Abell 754}

Abell 754 is another well studied galaxy cluster which shows
significant optical \cite{zz} and X-ray \cite{hb,hm} evidence of a major
merger event. This article will not go into details on the
observations of this cluster as they are presented elsewhere in these
proceedings \cite{ck} and in the literature \cite{kcecn}. The overall
picture \cite{mvmv} from optical and X-ray observations of the A754
system suggests that is undergoing a three body merger similar to
Abell 2256. Low frequency radio observations of the cluster reveal a
diffuse radio halo slightly offset from the cluster core, and possibly
two steep spectrum radio relics in the cluster periphery located on
either side of the cluster core. One of these relics (the eastern
relic) lies on the ram pressure flattened edge of the X-ray bar
suggesting a connection with the on-going merger event.

\section{Summary}

Both Abell 2256 and Abell 754 display very complex radio
emission. Although radio halos and relics are still relatively rare,
both of these are found in the apparent three body merger systems
A2256 and A754. It is becoming increasingly evident that the diffuse
radio emission in galaxy clusters is indeed related to major cluster
merger events. Combining optical, high resolution X-ray, and radio
observations of cluster systems can provide significant insight into
the merger history of the clusters. The radio and X-ray data also
place constraints on the relative dynamical roles of thermal and
nonthermal plasma components of the intracluster medium.

\section*{Acknowledgments}

We wish to thank our collaborators A.\ Cohen, N.\ Kassim, and D.\
Neumann. We also with to sincerely thank the organizers of this
meeting.

\end{document}